\documentclass[12pt,preprint]{aastex}
\usepackage{color}
\shorttitle{A BOUND-VIOLATING GROUP}
\shortauthors{Lee et al.}
\begin{document}
\title{A BOUND VIOLATION ON THE GALAXY GROUP SCALE: THE TURN-AROUND RADIUS OF NGC 5353/4}
\author{Jounghun Lee\altaffilmark{1}, 
Suk Kim\altaffilmark{2}, and Soo-Chang Rey\altaffilmark{2}}
\altaffiltext{1}{Astronomy Program, Department of Physics and Astronomy, Seoul National University, 
Seoul 151-742, Korea
\email{ jounghun@astro.snu.ac.kr}}
\altaffiltext{2}{Department of Astronomy and Space Science, Chungnam National University,
Daejeon 305-764, Korea}
\begin{abstract} 
The first observational evidence for the violation of the maximum turn-around radius on the galaxy group scale is presented.  
The NGC 5353/4 group is chosen as an ideal target for our investigation of the bound-violation because of its proximity, low-density environment, 
optimal mass scale, and existence of a nearby thin straight filament.  Using the observational data on the line-of-sight velocities and three-dimensional 
distances of the filament galaxies located 
in the bound zone of the NGC 5353/4 group, we construct their radial velocity profile as a function of separation distance from the group 
center and then compare it to the analytic formula obtained empirically by \citet{falco-etal14} to find the best-fit value of an adjustable 
parameter with the help of the maximum likelihood method.  
The turn-around radius of NGC 5353/4 is determined to be the separation distance where the adjusted analytic formula for the radial velocity 
profile yields zero. The estimated turn-around radius of NGC 5353/4 turns out to substantially exceed  the upper limit predicted by the spherical 
model based on the $\Lambda$CDM cosmology. Even when the restrictive condition of spherical symmetry is released, the estimated value 
is found to be only marginally consistent with the $\Lambda$CDM expectation. 
\end{abstract}
\section{INTRODUCTION}\label{sec:intro}

After the epoch of recombination, the growth of an overdense region in the primordial matter density field would be driven by 
the competition between its self-gravity and the cosmic expansion. The simplest approach to understanding the gravitational evolution 
of an initially overdense region is to regard it as a bound mini-universe with positive curvature and then to solve the Friedmann 
equations under the assumption that the region has a uniform density and a spherically symmetric shape \citep{GG72}. 
The Friedmann equations have a well known solution for this case, according to which the radius of the spherical region varies with time as a 
cycloid.  The maximum radius that the spherical overdense region with uniform density will reach during its cycloidal change is called the 
"turn-around radius," after which it begins to shrink back since the effect of its self-gravity becomes stronger than that of the cosmic 
expansion. In the subsequent nonlinear evolution, the region will go through relaxation, virialization and the eventual 
formation of a bound object \citep{lynden-bell67,white76,quinn-etal86,frenk-etal88}.
  
The virial radius of a bound object formed at the site of an initial overdense region develops only weak dependence on the background 
cosmology since it is determined mainly by the total mass contained in the region \citep{eke-etal96,KS96}. 
Whereas, the turn-around radius of an initial overdense region depends sensitively on the background cosmology as well as on the 
total mass, since at the turn-around moment the pulling effect of its self gravity is cancelled out by the pushing effect 
of dark energy on the region \citep[e.g.,][]{lahav-etal91,BS93,eke-etal96,eingorn-etal13,pavlidou-etal14}.  
For a background universe that has a larger amount of dark energy with a more negative equation of state, the turn-around radii of 
overdense regions would become smaller since their self-gravity becomes comparable earlier to the anti-gravity of dark 
energy \citep{PT14}. Hence, if the turn-around radii of the initial overdense regions corresponding to the galaxy groups and clusters could be 
estimated with high accuracy, they must provide a powerful independent diagnostics for the density and equation of state of dark 
energy \citep{pavlidou-etal14}. 

By using an analytic argument, \citet{PT14} recently put a stringent upper bound, $r_{\rm t, max}$, on the turn-around radius that a bound 
object with mass $M$ can have in a flat $\Lambda$CDM (cosmological constant $\Lambda$ and Cold Dark Matter) universe:  
\begin{equation}
\label{eqn:rt_max}
r_{\rm t, max} = {\cal{A}}\left(\frac{ 3MG}{\Lambda c^{2}}\right)^{1/3}\, ,
\end{equation}
where $\cal{A}$ is a proportionality factor whose value depends on the geometrical shape of an initial overdense region from which the 
bound object originates. Only, provided that the shape of an overdense region possesses a perfect spherical symmetry, 
the factor $\cal{A}$ equals unity, being independent of $M$. For the realistic case of a non-spherical region whose gravitational 
growth tends to proceed quite anisotropically \citep[e.g.,][]{zel70,BM96}, $\cal{A}$ slightly exceeds unity (roughly $1.3$) 
taking on the mass dependence \citep[see Figure 3 in][]{PT14}. From here on, the {\it spherical bound} ({\it non-spherical bound}) refers to 
$r_{\rm t, max}$ with ${\cal A}=1$ (${\cal A}>1$) in Equation (\ref{eqn:rt_max}). 

\citet{PT14} suggested that the "zero-velocity surface" around a bound object should be a good approximation of the true turn-around 
radius of the initial overdense site at which the object formed. Here the zero-velocity surface around a bound object is defined 
as the radial distance from the object center at which the peculiar motions generated by the object's gravity become equal in magnitude to 
the Hubble expansion. \citet{PT14} claimed that a violation of the spherical (non-spherical) bound given in Equation (\ref{eqn:rt_max}) 
by any bound object observed in the universe would contest the standard $\Lambda$CDM cosmology as a smoldering (smoking) 
gun counter-evidence, urging exploration of such bound-violating objects on the galaxy group and cluster scales.  
Unfortunately, however, the observational estimate of the turn-around radius of a galaxy group/cluster requires us to 
fulfill a tough mission: accurately measuring the peculiar velocities of the neighbor galaxies located in the bound-zone 
of the galaxy group/cluster (see Section \ref{sec:prediction} for the definition of a bound-zone) . 

Here, we develop a novel methodology to estimate the turn-around radius of a galaxy group/cluster without directly measuring 
the peculiar velocities.  It utilizes the universal formula for the radial velocity profile empirically derived by \citet{falco-etal14} 
to estimate the turn-around distances where the total radial velocities of the bound-zone galaxies vanish. We apply this new methodology 
to the NGC 5353/4 group, a galaxy group with mass $3\times 10^{13}\, M_{\odot}$, at a distance of $34.7$ Mpc from us, appeared to be 
surrounded by the low-density environment \citep{tully15}. 
Throughout this paper, to be consistent with \citet{PT14}, we assume the standard Planck model with 
$\Omega_{\rm m}=0.315,\ \Omega_{\Lambda}=0.685,\ h=0.673$ \citep{planck13} whenever the background cosmology has to be specified.

\section{THEORETICAL PREDICTION}
\label{sec:prediction}

The regions outside of the virial radius, $r_{\rm v}$, of an isolated massive object in the expanding universe can be divided into three 
distinct zones. The nearest to the object is the infall zone where the object's gravity defeats the cosmic expansion. If a lower-mass 
object is located in this zone, it is expected to undergo infall into the potential well of the massive object to eventually become its satellite 
\citep{zu-etal14}.
The farthest from the object is the Hubble zone where the cosmic expansion wins over the object's gravity.  If a smaller object is located in 
this zone, then its effective motion would be just the Hubble expansion since the gravitational influence from the massive object should be 
negligible. 
The in-between is the bound zone where the object's gravity is not dominant but strong enough to slow down the Hubble flow.  
The three zones (the infall, the bound, and the Hubble zones) correspond roughly to the following ranges of the separation distances, $r$, from 
the object center:  $r\le 2r_{\rm v}$, $3r_{\rm v}\le r < 10r_{\rm v}$ and $r > 10r_{\rm v}$, respectively \citep{PT14}. 

\citet{falco-etal14} has discovered the existence of the following universal profile of the bound-zone peculiar velocities:
\begin{equation}
\label{eqn:vp}
{\bf v}_{p}\cdot\hat{\bf r} = - V_{\rm c}\left(\frac{r}{r_{\rm v}}\right)^{-n_{\rm v}}\, .
\end{equation}
Here ${\bf v}_{p}$ is the peculiar velocity of a test particle at a separation distance of $r$ from a massive object with virial mass 
$M_{\rm v}$, $\hat{\bf r}$ is the unit vector in the radial direction from the center of the object to the position of a test particle, 
$r_{\rm v}$ is the virial radius of the object related to its virial mass as $M_{\rm v} = 4\pi\Delta_{c}r^{3}_{v}/3$ 
where $\Delta_{c}$ is $93.7$ times the critical density $\rho_{\rm crit,0}$, and $V_{\rm c}$ is the central velocity of a test 
particle at $r_{\rm v}$ given as $V_{\rm c}\approx \left(GM_{\rm v}/r_{\rm v}\right)^{1/2}$. The negative sign in the right-hand side (RHS) 
of Equation (\ref{eqn:vp}) reflects that the object's gravity is in the direction of $-\hat{\bf r}$. 

The universality of the above profile is manifested by the independence of the power-law index $n_{\rm v}$ from the masses and 
redshifts of the objects. Analyzing the numerical data around the cluster halos identified  from high-resolution $N$-body simulations for a flat 
$\Lambda$CDM cosmology, \citet{falco-etal14} have demonstrated that  Equation (\ref{eqn:vp}) with a constant power-law index of 
$n_{\rm v}\approx 0.42$ approximates the peculiar velocity profile of dark matter particles in the bound zones well, no matter what masses 
and redshifts the cluster halos have. Their numerical result implied that the expectation value of the peculiar velocity at any point of the 
bound zone around a cluster could be theoretically evaluated.

It has also been found by \citet{falco-etal14} that the matter-to-halo bias does not alter the functional form of Equation (\ref{eqn:vp}) by 
demonstrating that it validly approximates not only the peculiar velocity profile of dark matter particles, but also that of the galactic halos 
located in the bound zone. This numerical finding has elevated the practicality of Equation (\ref{eqn:vp}) since what is directly 
observable is not the positions of dark matter particles but those of the bound-zone galaxies.
Here, we claim that the very existence of the universal peculiar velocity profile in the bound zone of a massive object 
(like a galaxy group and cluster) should enable us to estimate its turn-around radius without directly measuring the peculiar velocity field 
in the bound-zone.  

Note that the bound-zone of a galaxy cluster is similar to the linear regime when the initial proto-cluster evolved until the turn-around moment. 
Accordingly, the peculiar velocities of the bound-zone galaxies predicted by Equation (\ref{eqn:vp}) can be treated as the linear perturbations to 
the zero mean and thus must be a good approximation to the peculiar velocity of the initial proto-cluster region.  Now, extrapolating 
Equation (\ref{eqn:vp}) to the turn-around moment, we determine the value of $r_{\rm t}$ as the separation distance at which the following 
equality holds true. 
\begin{equation}
\label{eqn:rt}
H_{\rm 0}r_{\rm t} = V_{\rm c}\left(\frac{r_{\rm t}}{r_{\rm v}}\right)^{-n_{\rm v}}\, .
\end{equation}
The left-hand side (LHS) of Equation (\ref{eqn:rt}) represents nothing but the Hubble expansion while the RHS is the peculiar velocity of an initial 
proto-cluster region in the radial direction at the turn-around radius $r_{\rm t}$ approximated by the predictable peculiar velocity of a bound-zone 
galaxy in the radial direction at the same distance $r_{\rm t}$ . 

To solve Equation (\ref{eqn:rt}) in practice, we consider the total radial velocity profile, $v_{r}(r)$, defined as the combination of 
the Hubble expansion with the peculiar velocity in the radial direction: 
\begin{equation}
\label{eqn:vr}
v_{r}(r) = H_{\rm 0}r - V_{\rm c}\left(\frac{r}{r_{\rm v}}\right)^{-n_{\rm v}}\, ,
\end{equation}
and then look for the range of $r$ where the function, $v_{r}(r)$, crosses the zero line.  It should be emphasized that 
finding a solution to $v_{r}(r_{\rm t})=0$ with Equation (\ref{eqn:vr}) requires information only on the radial positions of the bound-zone galaxies 
but not on their peculiar velocities. 

\begin{figure}[b]
\begin{center}
\epsscale{1.0}
\plotone{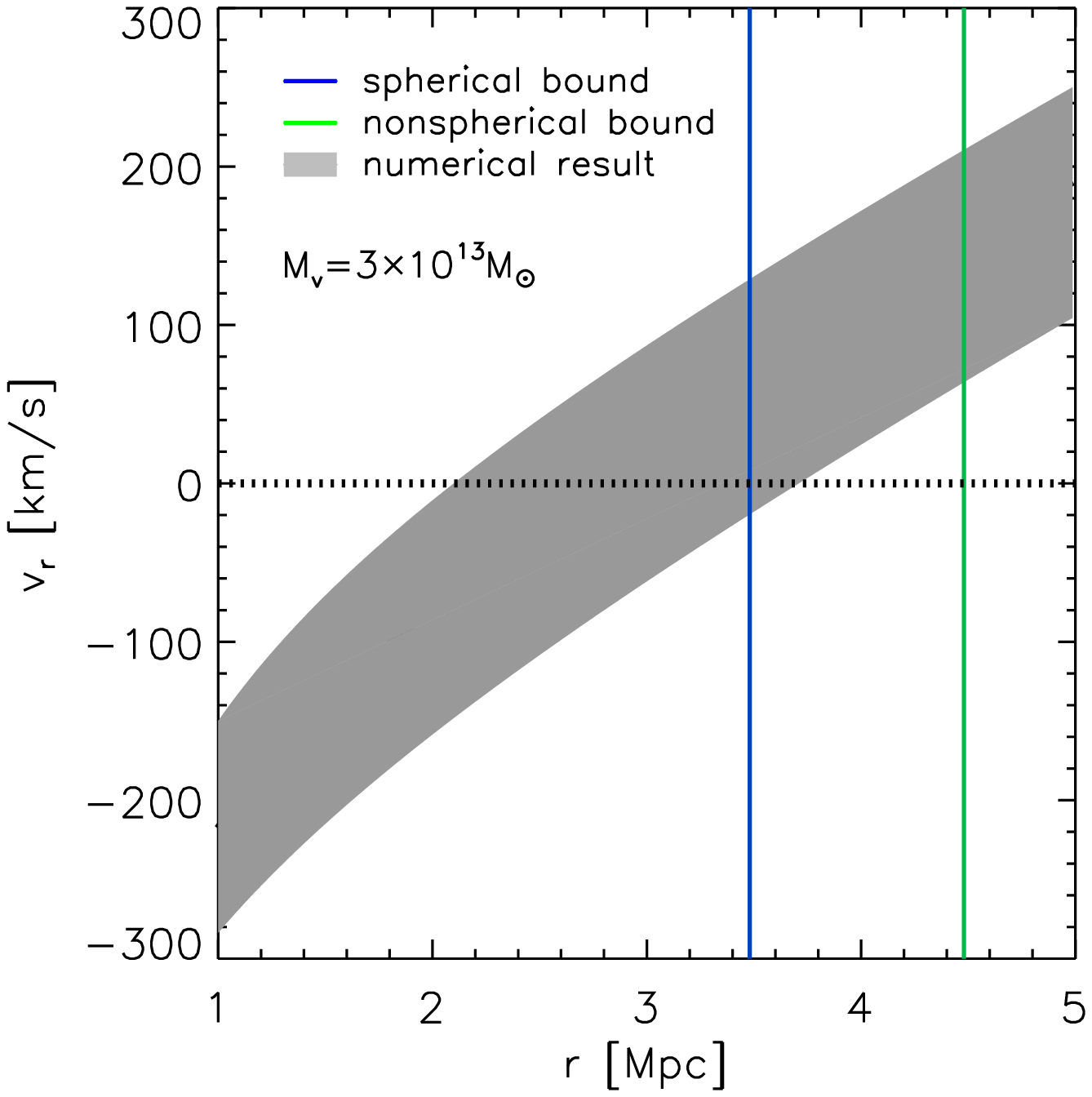}
\caption{Radial velocity profile of the galaxies ($v_{r}$) located in the bound zone of a galaxy group with virial mass 
$M_{\rm v}=3\times 10^{13}\,M_{\odot}$ (gray region) determined by Equation (\ref{eqn:vr}). The turn-around radius ($r_{\rm t}$) of the galaxy 
group is in the range of the separation distances between the bound-zone galaxies and the group center, $r$, where the gray region 
touches the dotted-line ($v_{\rm r}=0$). The blue and green solid lines indicate the locations of the spherical and the non-spherical bounds 
predicted by the standard $\Lambda$CDM cosmology \citep{PT14}.}
\label{fig:vr}
\end{center}
\end{figure}

We plot Equation (\ref{eqn:vr}) for the case of $M_{\rm v}=3\times 10^{13}\, M_{\odot}$ (the virial mass of NGC 5353/4, see \citet{tully15}) 
as the gray region in Figure \ref{fig:vr}.  Given the numerical result of $n_{\rm v} = 0.42\pm 0.16$ and $V_{c}=(0.8\pm 0.2)(GM_{\rm v}/r_{\rm v})^{1/2}$ 
obtained by \citet{falco-etal14}, we let the power-law index $n_{\rm v}$ and the multiplicative constant 
$a\equiv V_{\rm c}(GM_{\rm v}/r_{\rm v})^{-1/2}$ vary in the ranges of $2.6\le n_{\rm v}\le 5.8$ and $0.6\le a\le 1.0$, respectively, to 
plot Equation (\ref{eqn:vr}). 
The section of the dotted line ($v_{r}=0$) inside the gray region represents the expected range of the turn-around radius of a galaxy group 
with $M_{\rm v}=3\times 10^{13}\, M_{\odot}$ for a flat $\Lambda$CDM cosmology.  The blue and green solid lines 
indicate the locations of the spherical and non-spherical bounds, respectively. 
The comparison of the estimated range of $r_{\rm t}$ with the spherical and non-spherical bounds in Figure \ref{fig:vr} leads us to 
answer the question of whether it is possible to find a bound-violating galaxy group with mass $M_{\rm v}=3\times 10^{13}\, M_{\odot}$ in a flat 
$\Lambda$CDM universe. As can be seen, the crossing between the blue (green) solid and the black dotted lines occurs inside (outside) the gray region. 
This result indicates that although it is not impossible for such a galaxy group to violate the spherical bound in a flat $\Lambda$CDM model, 
the violation of the non-spherical bound would rarely occur on that mass scale.

\begin{figure}
\begin{center}
\epsscale{1.0}
\plotone{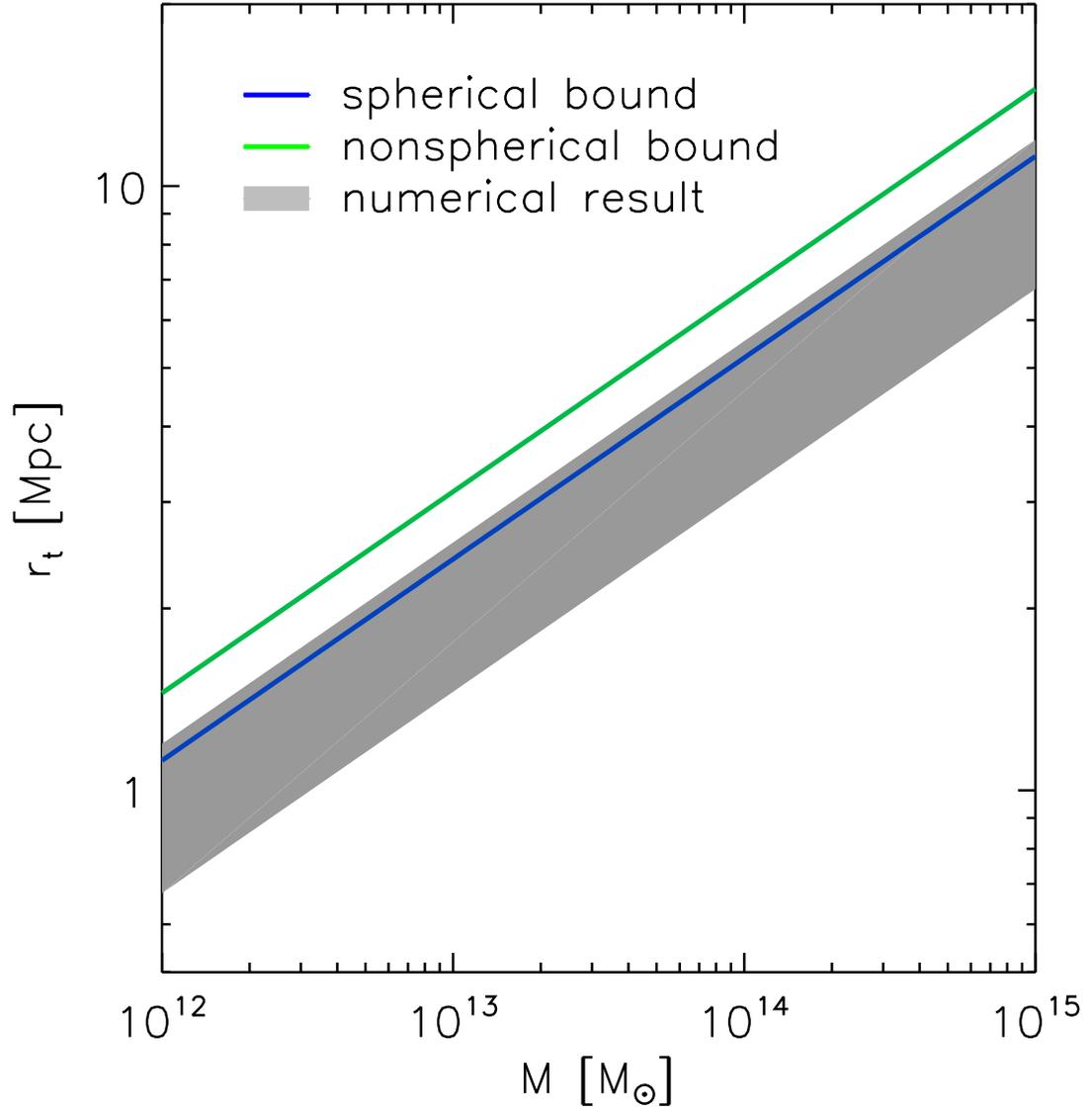}
\caption{Variation of the turn-around radius ($r_{\rm t}$) with the virial mass of a bound object (gray region) estimated 
by solving Equation (\ref{eqn:rt}). 
The spherical and non-spherical upper bounds, $r_{\rm t, max}$ predicted by the $\Lambda$CDM model are plotted 
as blue and green solid lines, respectively.}
\label{fig:rt_m}
\end{center}
\end{figure}

To see whether this critical prediction depends on the mass scale,  we vary the values of $M_{\rm v}$ in Equation (\ref{eqn:vr}) and find the 
turn-around radius $r_{\rm t}$ as a function of $M_{\rm v}$, which is shown in Figure \ref{fig:rt_m} as gray region. The blue and green solid lines 
correspond to the spherical and non-spherical bounds as a function of $M_{\rm v}$, respectively. As can be seen, the blue (green) solid line is 
inside (outside) the gray region at all mass scales.  Thus, the $\Lambda$CDM prediction against the bound-violation is extended to all mass scales:  
it is quite unlikely to find a bound object whose turn-around radius exceeds the non-spherical bound in the standard flat $\Lambda$CDM cosmology, 
no matter what mass scale the object has.  
 
 \section{NGC 5353/4: A BOUND-VIOLATING STRUCTURE}
\subsection{Radial Velocity Profile of the Filament Galaxies around NGC 5353/4}\label{sec:ngc5354}

\citet{PT14} suggested that the optimal target for the investigation of the bound violation should be a "nearby galaxy group located in the 
low-density environment." 
The proximity condition is necessary to minimize the observational uncertainties.  A low-density environment around a target is required to 
minimize the difference between the virial and the turn-around masses. Note that $M$ in Equation (\ref{eqn:rt_max}) represents the 
turn-around mass enclosed by the turn-around radius $r_{\rm t}$ which is expected to be larger than the virial mass $M_{\rm v}$ 
\citep{PT14}.  The turn-around mass is often approximated by the virial mass since the former is unmeasurable while a variety of methods 
has been developed to measure the latter. This approximation, however, works well for the case that a given target is located in the low-density 
environment. 

The reason for the preference of the groups to the clusters lies in the hierarchical nature of structure formation process: the galaxy 
groups are more relaxed systems than the galaxy clusters since the former must have formed earlier than the latter. 
Thus, the systematic errors produced by the deviation of the true dynamical state from complete relaxation would contaminate 
the measurements of the virial masses of the galaxy groups less severely than those of the galaxy clusters. 

Here, we require one more condition in addition to the above three for an optimal target: the presence of a thin straight filament 
in the bound-zone. According to \citet{falco-etal14}, the total radial velocities and three-dimensional positions of the galaxies could be 
readily inferred from the observable line-of-sight velocities and the two-dimensional projected positions, if the bound-zone galaxies 
are located along one-dimensional filament.  If the galaxies are distributed along one-dimensional filaments, they exhibit coherent 
motions along the filaments, which in turn makes it easier to judge whether or not the observed galaxies are located in the bound-zone 
of a target (see also S. Kim et al. 2015, in preparation).
In our previous work \citep{lee-etal15}, which reconstructed the radial velocity profile of the Virgo cluster and compared it to Equation 
(\ref{eqn:vr}), it was confirmed that the presence of a thin straight filament is a key ingredient for the accurate reconstruction of the radial 
velocity profile of the bound-zone galaxies. 

We find  that the NGC 5353/4 group meets all of the above four conditions. It is only $34.7$ Mpc away from us and observed to be located in  
the low-density environment \citep{TT08, tully15}. 
The three-dimensional position of the center of the NGC 5353/4 group (say, ${\bf x}_{c}$) is measured by using available 
information on its equatorial coordinates and comoving distance.  To estimate the virial mass, $M_{\rm v}$, of NGC 5353/4, \citet{TT08} used 
two distinct methods: one was based on the measurements of the velocity dispersions of the NGC 5353/4 satellites and 
the other employed the projected mass estimator given by \citet{heisler85}. They took the average over the two estimates to find 
$M_{\rm v}=2.1\times 10^{13}\, M_{\odot}$. Recently, however, \citet{tully15} has updated $M_{\rm v}$ to a slightly higher value, 
$3\times 10^{13}\, M_{\odot}$. The corresponding virial radius of NGC 5353/4 is determined to be  $r_{\rm v}=1.1$ Mpc
by using the relation of $r_{\rm v} = \left[3M_{\rm v}/(4\pi \Delta_{c})\right]^{1/3}$.  Adopting the conservative definition of 
\citet{falco-etal14},  we confine the bound-zone of the NGC 5353/4 group to the region enclosed by a spherical shell whose inner 
and outer radii equal $3r_{\rm v}$ and $8r_{\rm v}$, respectively. 

S. Kim et al. (2015, in preparation) recently detected around the NGC 5353/4 group a thin straight filament, which is mainly composed of 
dwarf galaxies with B-band magnitudes $12.83\le m_{\rm B}\le 19.54$ in the redshift range of $0.007\le z\le 0.011$. 
Noting that the filament is elongated along the direction toward the NGC 5353/4 group and analyzing the line-of-sight velocities of the filament 
galaxies, S. Kim et al. (2015, in preparation) have concluded that the filament is located in the bound zone of the NGC 5353/4 group. 
Since the majority of the filament galaxies has been found to be the dwarf galaxies, the most gravitational influence that the filament galaxies 
in the bound zone would feel should come from the NGC 5353/4 group. Information on the comoving distances of $17$ member galaxies of the 
filament are found available at  the NASA/IPAC Extragalactic Database\footnote{https://ned.ipac.caltech.edu}.  

The three-dimensional positions, ${\bf x}$, of the $17$ filament galaxies are determined and their separation displacement vectors ${\bf r}$ 
from the group center are calculated as ${\bf r}\equiv {\bf x}-{\bf x}_{c}$.  Then, we select only those among the $17$ filament galaxies 
whose separation distances, $r\equiv \vert{\bf r}\vert$, satisfy the bound-zone condition of $3\le r/r_{\rm v}\le 8$. From here on, the filament 
galaxies, which have their comoving distances measured and belong to the bound-zone of NGC 5353/4, are referred to as {\it the bound-zone 
filament galaxies} of NGC 5353/4. A total of four bound-zone filament galaxies is selected.

The angle, $\beta$, at which the radial direction, $\hat{\bf r}\equiv {\bf r}/r$, of each bound-zone filament galaxy is inclined to the line-of-sight 
direction of the group center, $\hat{\bf x}_{c}\equiv {\bf x}_{c}/x_{c}$, can be determined as $\cos\beta =\hat{\bf r}\cdot\hat{\bf x}_{c}$. 
Now, the radial velocity of each bound-zone filament galaxy in unita of km s$^{-1}$ can be evaluated from the measurable inclination angle $\beta$ 
and the redshift difference $\Delta z$ between the center of the NGC 5353/4 and its bound-zone filament galaxies, 
\begin{equation}
\label{eqn:vlos}
v_{r}(r) = \frac{c\Delta z}{\cos\beta}\, ,
\end{equation}
where $c\Delta z$ is basically the line-of-sight velocity of a bound-zone filament galaxy relative to the NGC 5353/4 center.

\subsection{Testing the $\Lambda$CDM Cosmology with NGC 5353/4}\label{sec:test}

In Section \ref{sec:ngc5354} we have measured the radial velocities of the bound-zone filament galaxies of NGC 5353/4 from observations.  
We are now ready to adjust Equation (\ref{eqn:vr}) to this observational result and to eventually estimate the turn-around radius of NGC 5353/4 
by equating the RHS of Equation (\ref{eqn:vr}) to zero. The adjustable parameter is nothing but the power-law index, $n_{\rm v}$, in the RHS of 
Equation (\ref{eqn:vr}).
As mentioned in Section \ref{sec:prediction}, \citet{falco-etal14} found $n_{\rm v}=0.42\pm 0.16$ from their numerical experiment for the standard flat 
$\Lambda$CDM cosmology. Without ruling out the possibility that the true universe deviates from the standard flat $\Lambda$CDM cosmology, 
it is not unreasonable for us to suspect that the value of $n_{\rm v}$ might deviate from the estimate of \citet{falco-etal14}. 
Moreover, our previous work has found that, although the functional form of Equation (\ref{eqn:vr}) itself describes well the 
reconstructed radial velocity profile of the Virgo cluster from observational data, the best agreement is reached when the power-law 
index $n_{\rm v}$ has a lower (negative) value than $0.42$ \citep{lee-etal15}. 

Varying the power-law index of $n_{\rm v}$, we fit the RHS of Equation (\ref{eqn:vr}) to the measured radial velocities of the bound-zone 
filament galaxies from information on $\beta$ and $c\Delta z$ (see Eq.(\ref{eqn:vlos})). 
Then, we search for the best-fit value of $n_{\rm v}$ that maximizes the following likelihood function:
\begin{equation}
\label{eqn:like1}
p(n_{\rm v}\vert M_{\rm v}, r_{i}, z_{i}, \beta_{i}) \propto \exp\left(-\frac{\chi^{2}_{\nu}}{2}\right)\, , 
\end{equation}
where $\chi^{2}$ is the reduced chi-square given as
\begin{equation}
\label{eqn:like2}
\chi^{2}_{\nu}(n_{\rm v}\vert M_{\rm v}, r_{i}, z_{i}, \beta_{i}) = \frac{1}{\nu}\sum_{i=1}^{N_{\rm g}}
\left[v^{\rm ob}(z_{i},\beta_{i}) - v^{\rm th}(r_{i}; M_{\rm v}, n_{\rm v})\right]^{2}\, .
\end{equation}
Here $N_{\rm g}$ denotes the number of the bound-zone filament galaxies, $v^{\rm ob}(r_{i})\equiv c\Delta z_{i}/\cos\beta_{i}$ 
represents the observational radial velocity of the $i$th bound-zone filament galaxy with redshift difference $\Delta z_{i}$ and inclination 
angle $\beta_{i}$ estimated  by Equation (\ref{eqn:vlos}),  and $v^{\rm th}(r_{i}; n_{\rm v})$ is the theoretical radial velocity at a separation 
distance $r_{i}$ predicted by Equation (\ref{eqn:vr}). Note that the degree of freedom $\nu$ equals $N_{\rm g}-1$ since Equation (\ref{eqn:vr}) 
has only one adjustable parameter, $n_{\rm v}$. 

\begin{figure}
\begin{center}
\epsscale{1.0}
\plotone{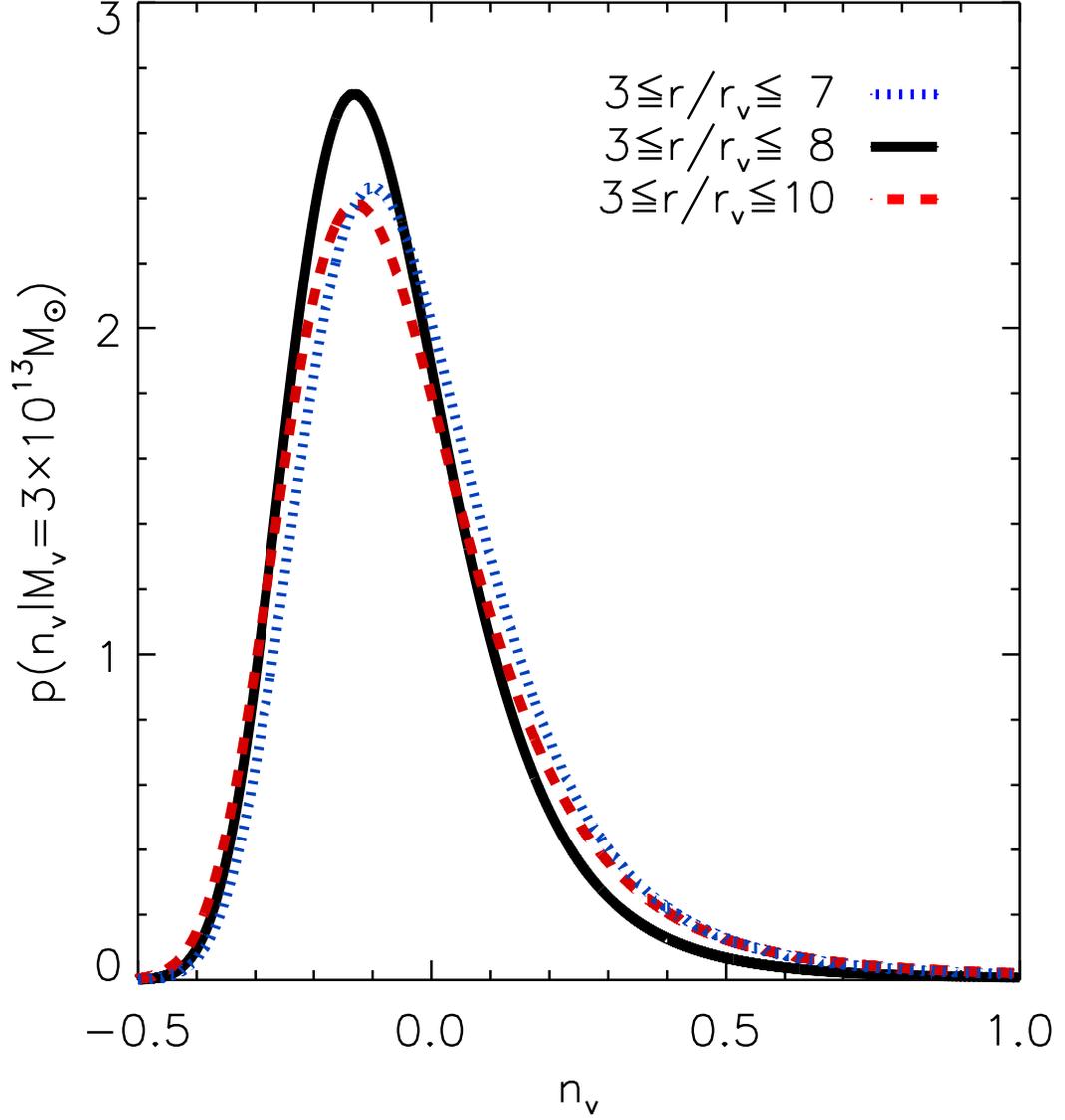}
\caption{Posterior probability density distributions of the power-law index $n_{\rm v}$ in the analytic formula of \citet{falco-etal14}  
fitted to the observed radial velocities of the bound-zone filament galaxies around NGC 5353/4. The blue dotted, black solid, and 
red dashed lines correspond to the cases in which the separation distances $r$ of the bound-zone filament galaxies from NGC 5353/4 are 
in the range of $3\le r/r_{\rm v}\le 7$, $3\le r/r_{\rm v}\le 8$, and $3\le r/r_{\rm v}\le 10$, respectively, where $r_{\rm v}$ is the virial 
radius of NGC 5353/4.}
\label{fig:p_nv}
\end{center}
\end{figure}

Normalizing $p(n_{\rm v})$ to  satisfy $\int dn_{\rm v}\,p(n_{\rm v}) = 1$, we plot it as a black solid line in Figure \ref{fig:p_nv}. 
As can be seen, the likelihood function $p(n_{\rm v})$ reaches its sharp peak at $n_{\rm v,p}=-0.13$. The evaluation of the upper and 
lower errors (say $\sigma_{\rm u}$ and $\sigma_{\rm l}$) associated with the best-fit value $n_{\rm v,p}$ as 
\begin{equation}
\label{eqn:sigu}
\int_{n_{\rm v,p}}^{n_{\rm v,p}+\sigma_{\rm u}}\, dn_{\rm v}^{\prime}\, p(n_{\rm v}^{\prime} \vert M_{\rm v}, r_{i}, z_{i}, \beta_{i})=
\int^{n_{\rm v,p}}_{n_{\rm v,p}-\sigma_{\rm l}}\, dn_{\rm v}^{\prime}\, p(n_{\rm v}^{\prime} \vert M_{\rm v}, r_{i}, z_{i}, \beta_{i}) = 0.34
\end{equation}
yields $\sigma_{\rm u}=0.14$ and $\sigma_{\rm l}=0.16$,  respectively. 

\begin{figure}
\begin{center}
\epsscale{1,0}
\plotone{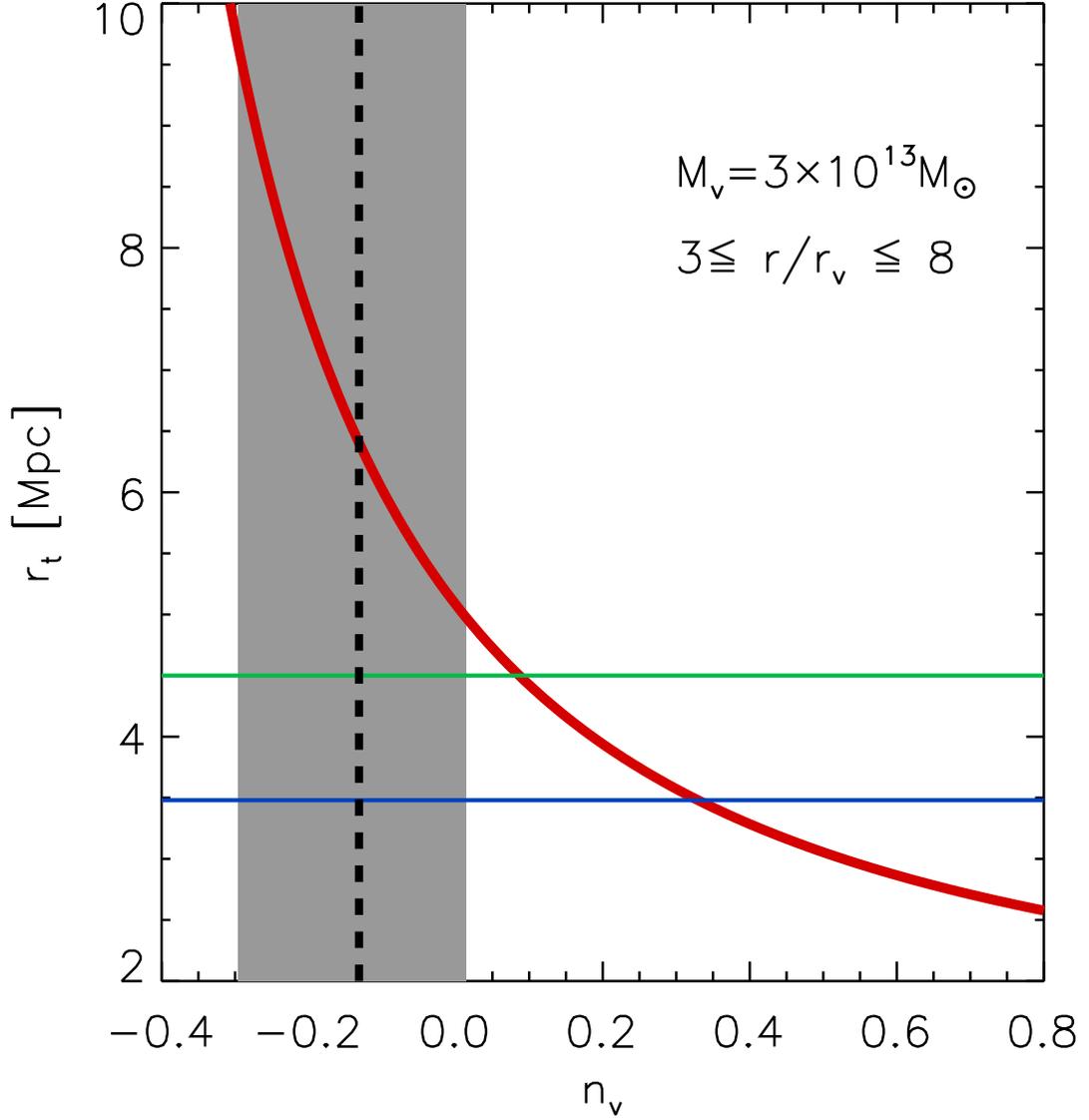}
\caption{Variation of the turn-around radius with the power-law index $n_{\rm v}$ of the radial velocity profile 
in the bound zone of a galaxy group with mass $M_{\rm v}=3\times 10^{13}\, M_{\odot}$ (red solid line). 
The dashed line and the gray region indicate the location of the best-fit $n_{\rm v}$ and the width of the associated $1\sigma$ scatter 
around the best-fit $n_{\rm v}$, respectively, which are estimated by fitting the analytic formula of \citet{falco-etal14} to the observed 
radial velocities of the bound-zone filament galaxies around NGC 5353/4 where the bound-zone range is given as 
$3\le r/r_{\rm v}\le 8$. The blue and green solid lines indicate the locations of the spherical and non-spherical bounds 
(i.e., the upper limit of the turn-around radius, $r_{\rm t, max}$,  predicted by the standard $\Lambda$CDM model, respectively. }
\label{fig:rt_nv8}
\end{center}
\end{figure}

Now that the best-fit power-law index, $n_{\rm v, p}$, for Equation (\ref{eqn:vr}) has been determined, the corresponding turn-around 
radius $r_{\rm t}$ can be calculated by equating Equation (\ref{eqn:vr}) to zero. Figure \ref{fig:rt_nv8} plots how the turn-around radius 
varies with the power-law index $n_{\rm v}$ as a red solid line.  The location of the best-fit value, $n_{\rm v,p}$, and the amount of the 
associated errors, $\sigma_{\rm u}$ and $\sigma_{\rm l}$, obtained by means of the maximum likelihood method are shown as the black 
dashed line and gray regions, respectively.  The blue and the green solid lines indicate the locations of the spherical and nonspherical 
bounds, respectively. 
Projecting the section of the red solid line overlapped with the gray region onto the vertical axis in Figure \ref{fig:rt_nv8} allows us 
to determine the range of the turn-around radius of NGC 5353/4. As can be seen, the best-fit value $r_{\rm t}$ is one $\sigma$ higher 
than the upper bound predicted by the $\Lambda$CDM model.  Although the signal of the bound-violation is not so statistically significant 
less than $2\sigma$, we believe that the NGC 5353/4 group is a strong candidate for the bound-violation on the group scale since its 
best-fit turn-around radius turns out to exceed not only the spherical, but also the non-spherical, upper bound. 

\begin{figure}[b]
\begin{center}
\epsscale{1.0}
\plotone{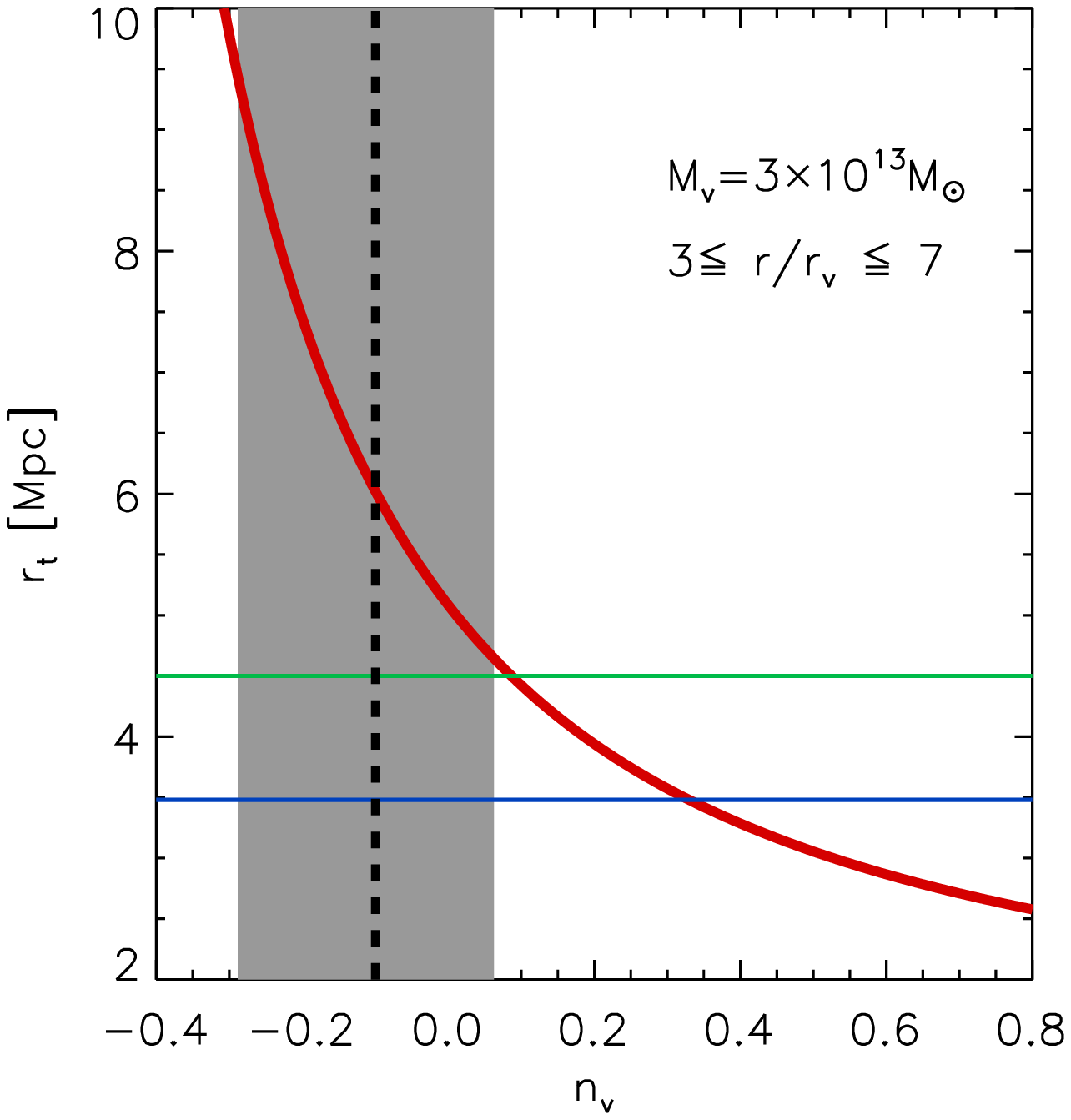}
\caption{Same as Figure \ref{fig:rt_nv8}, but for the case in which the bound-zone range is given as $3\le r/r_{\rm v}\le 7$. }
\label{fig:rt_nv7}
\end{center}
\end{figure}

\begin{figure}[ht]
\begin{center}
\epsscale{1.0}
\plotone{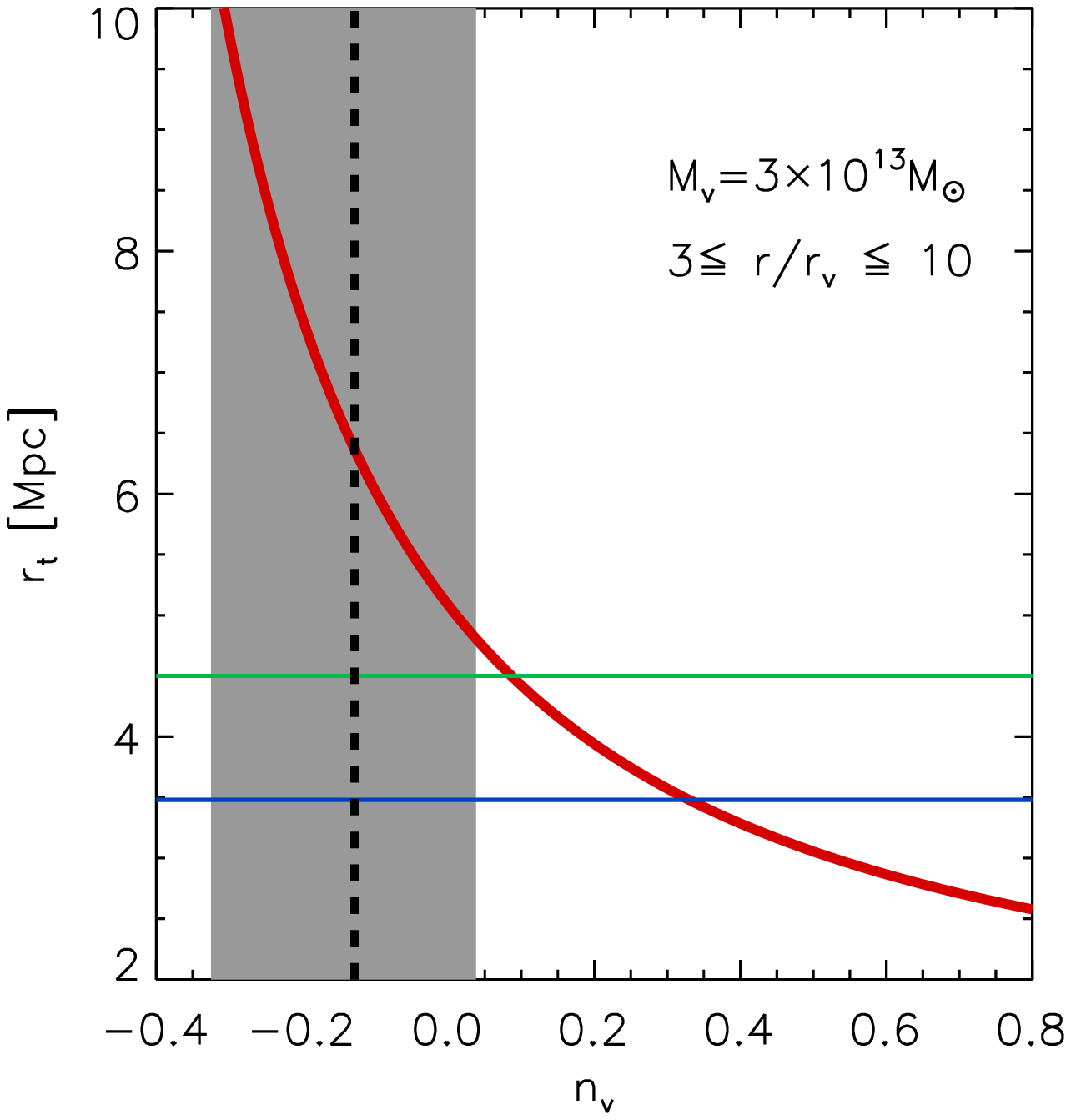}
\caption{Same as Figure \ref{fig:rt_nv8}, but for the case in which the bound-zone range is given as $3\le r/r_{\rm v}\le 10$. } 
\label{fig:rt_nv10}
\end{center}
\end{figure}

We also examine how robust the fitting result is against the change of the bound-zone range from 
$3\le r/r_{\rm v}\le 8$ into $3\le r/r_{\rm v}\le 7$ and into $3\le r/r_{\rm v}\le 10$. 
The likelihood functions,  $p(n_{\rm v})$ for these two cases of the bound zone range are shown as red dashed and blue dotted lines, 
respectively, in Figure \ref{fig:p_nv}. As can be seen, the change of the bound-zone range produces a thicker tail of 
$p(n_{\rm v})$ in the high $n_{\rm v}$ section and moves the location of the peak, $n_{\rm v,p}$, to a slightly higher value. 
Figures \ref{fig:rt_nv7} and \ref{fig:rt_nv10} plot the same as Figure \ref{fig:rt_nv8} but for the cases of 
$3\le r/r_{\rm v}\le 7$ and $3\le r/r_{\rm v}\le 10$, respectively, which demonstrate that the change of the bound-zone range enlarges 
the errors of $r_{\rm t}$, but does not alleviate the bound-violation by the NGC 5353/4 group much. 
Table \ref{tab:nv} lists the numbers of the bound-zone filament galaxies, the corresponding best-fit power-law index and the 
resulting ranges of the turn-around radius $r_{\rm t}$ for three different cases of the bound-zone limit.  

\section{SUMMARY AND DISCUSSION}
\label{sec:con}

Our work was inspired and urged by the seminal paper of \citet{PT14} which theoretically proved the existence of the maximum 
turn-around radii, $r_{\rm t, max}$, of  massive objects and proposed it as an independent test of the $\Lambda$CDM model by 
showing its sensitive dependence on the density and equation of state of dark energy \citep[see also][]{pavlidou-etal14}.  
In the current work, we have taken a step toward realizing the ingenious idea of \citet{PT14} by accomplishing the following.  First, we 
have devised a new methodology to infer the turn-around radii of galaxy groups/clusters without information on the peculiar velocity 
field.  Based on the numerical finding of  \citet{falco-etal14} that the radial velocity profile of dark matter particles in the bound zone of 
massive objects displays universality, this new methodology employs the extrapolation of the universal profile to the turn-around moment 
at which the radial velocities would vanish. 

Second, we have applied this new methodology to the nearby galaxy group, NGC 5353/4, around which a thin straight 
filament composed mainly of the dwarf galaxies has been detected (S. Kim et al. 2015, in preparation). Assuming that the radial velocity profile 
of the filament galaxies in the bound zone of NGC 5353/4 follows the universal form well, we have shown that 
having information on the spatial positions of the filament galaxies suffices to determine the best-fit value of the 
characteristic parameter of the profile.  Then, we have constrained the ranges of the turn-around radius, $r_{\rm t}$, of NGC 5353/4 by 
locating the separation distance at which the radial velocity profile with the best-fit parameter crosses the zero line. The comparison of the 
constrained range of $r_{\rm t}$ with the theoretical upper bound, $r_{\rm t, max}$, derived by \citet{PT14} has led us to detect 
a $2\sigma$ signal of the violation of the spherical bound and a $1.3\sigma$ signal of the violation of the non-spherical bound. 
Weak as the signal may appear at first sight, our result is significant because it reports the first case of possible violation of the 
non-spherical bound on the galaxy group scale.

All of the previously reported cases of the bound-violation occurred on the cluster or supercluster scales but none on the group scale 
\citep[see Figure 1 in][]{PT14}. Lack of evidence for the bound-violating cases on the galaxy group scale had been interpreted 
as an indication that the signals of the bound-violation detected on the cluster scale might be spurious ones generated by 
large uncertainties in the measurements of the virial masses of the clusters as well as the peculiar velocities of their bound-zone 
galaxies \citep{naso-etal11,kara-etal14}. 
Our result provides the first observational counter-evidence against the $\Lambda$CDM prediction for $r_{\rm t, max}$ on the galaxy 
group scale, which has two crucial implications. First, the observed signals of the bound-violation on the cluster scale should not be  
interpreted just as false ones but deserve thorough reinvestigation by making efforts to estimate the turn-around radii more accurately. 

Second, the turn-around radii of those galaxy groups, which were estimated and found to match the $\Lambda$CDM prediction well in previous 
studies \citep{KK06,kara-etal07} may need to be reestimated without placing too much confidence on the measurements of the peculiar 
velocities of their bound-zone galaxies. Since the galaxy groups have weaker self-gravity and a smaller number of bound-zone galaxies than 
the galaxy clusters, it is in fact delicately more difficult to infer their turn-around radii by directly measuring the peculiar velocities of the 
bound-zone galaxies. Our methodology has allowed us to overcome this difficulty, being capable of producing a robust result that is expected 
to be less severely contaminated by observational errors. Yet, as mentioned in \citet{PT14}, given the inherent stochastic nature of the 
structure formation process, it requires more counter-evidence on the galaxy group scale to contest the $\Lambda$CDM cosmology 
with the $r_{\rm t}$ values estimated by our methodology. 

We admit that our result may suffer from the poor-number statistics. Only four bound-zone filament galaxies have been 
used in our analysis to determine the best-fit value of the power-law index of the radial velocity profile around NGC 5353/4, simply because it is only 
those four galaxies that already have their comoving distances independently measured . It would definitely be desirable to measure the comoving 
distances of more bound-zone filament galaxies of NGC 5353/4 and then to reestimate the best-fit value of the power-law index of the radial velocity 
profile by using them all, which is, however, beyond the scope of this paper. 
We also admit that not all possible errors including hidden systematics have been taken into account to derive our results. For instance, if  
the uncertainties in the measurements of the virial mass of NGC 5353/4 and the separation distances to the bound-zone filament 
galaxies were known and accounted for,  the errors in the final estimate of $r_{\rm t}$ would become larger, which might in turn alleviate 
the tension with the $\Lambda$CDM prediction for $r_{\rm t, max}$. Furthermore, although the NGC 5353/4 group appears to be located 
in the low-density environment \citep{tully15}, non-negligible difference between the virial and the turn-around masses of NGC 5353/4 may 
exist and could produce other uncertainties regarding the estimate of its turn-around radius. 

Recently, \citet{faraoni15} studied how the deviation of gravitational law from the General Relativity would affect the value of the maximum 
turn-around radii and derived a general expression for $r_{\rm t, max}$ in modified gravity (MG) models from the first principles.  Our 
methodology will be useful to put an independent constraint on the MG models by efficiently estimating the turn-around radii of several 
nearby galaxy groups  and comparing the estimates with the MG predictions given by \citet{faraoni15}.  
However, some numerical work has to precede such a test with our methodology. Strictly speaking, the functional 
form of the radial velocity profile (Eq.(\ref{eqn:vr})) empirically drawn by \citet{falco-etal14} from $N$-body simulations is applicable only for a flat 
$\Lambda$CDM cosmology and it is not guaranteed that the same functional form can be applied for the cases of MG models. 
It will be necessary to investigate if and how not only the value of the power-law index but also the functional form itself of the radial 
velocity profile of the bound-zone galaxies differs from Equation (\ref{eqn:vr}) in MG models.  We plan to work on this project as well as on 
testing the robustness of the current result by using more improved data sets. We hope to report the result elsewhere in the near future.
 
\acknowledgments

This work was supported by a research grant from the National Research Foundation (NRF) of Korea to the Center for 
Galaxy Evolution Research  (NO. 2010-0027910).  J.L. also acknowledges the financial support of the Basic Science Research 
Program through the NRF of Korea funded by the Ministry of Education (NO. 2013004372).
S.C.R. acknowledges the support of the Basic Science Research Program through the NRF funded by the Ministry of Education, 
Science, and Technology (NRF-2015R1A2A2A01006828).
S.K. acknowledges support from the National Junior Research Fellowship of NRF (No. 2011-0012618).

\clearpage
\clearpage
\clearpage
\clearpage
\clearpage
\clearpage
\clearpage
\begin{deluxetable}{cccc}
\tablewidth{0pt}
\setlength{\tabcolsep}{5mm}
\tablecaption{Ratios of the separation distances ($r$) of the bound-zone filament galaxies to the virial radius 
($r_{\rm v}$) of NGC 5353/4, the number of the bound-zone filament galaxies ($N_{\rm g}$), and the best-fit value of the 
power-law index of the radial velocity profile ($n_{\rm v}$) and the estimated range of the turn-around radius of NGC 5353/4 
($r_{\rm t}$).}
\tablehead{$r/r_{\rm v}$ & $N_{\rm g}$ &  $n_{\rm v}$ & $r_{\rm t}$ \\ 
& & & (Mpc)}
\startdata
$[3\, , 7]$  & $3$ & $-0.10^{+0.16}_{-0.18}$ & $[4.67,\ 9.21]$\\
$[3\, , 8]$  & $4$ & $-0.13^{+0.15}_{-0.16}$ & $[4.93,\ 9.49]$\\
$[3\, ,10]$ & $5 $ & $-0.13^{+0.17}_{-0.20}$ & $[4.79,\ 10.79]$
\enddata
\label{tab:nv}
\end{deluxetable}
\end{document}